\documentclass[conference]{IEEEtran}
\usepackage{cite}

\ifCLASSINFOpdf
  \usepackage[pdftex]{graphicx}
\else
\fi
\usepackage{url}

\usepackage{fixltx2e}

\begin{document}
%
\title{868 MHz Wireless Sensor Network - A Study}

\author{\IEEEauthorblockN{Pushpam Aji John, Rudolf Agren, Yu-Jung Chen, Christian Rohner, and Edith Ngai}
\IEEEauthorblockA{Department of Information Technology\\
Uppsala University, Sweden\\
 }} 


%


\maketitle

\begin{abstract}

Today 2.4 GHz based wireless sensor networks are increasing at a tremendous pace, and are seen in widespread applications. Product innovation and support by many vendors in 2.4 GHz makes it a preferred choice, but the networks are prone to issues like interference, and range issues. On the other hand, the less popular 868 MHz in the ISM band has not seen significant usage. In this paper we explore the use of 868 MHz channel to implement a wireless sensor network, and study the efficacy of this channel.

\end{abstract}



%
\IEEEpeerreviewmaketitle

\section{Introduction}
Popularity of 2.4 GHz sensor networks have promulgated into all aspects of our life. Today it has been used in wide range of applications from monitoring body to improving efficiency in petroleum fields \cite{survey:overview}. The adoption of it has made the proliferation of standards such as IEEE 802.15.4, ZigBee, Wireless HART and 6LowPan \cite{survey:overview}. These standards innately promote low cost, low power, and larger deployments. 

In addition, the proliferation of WiFi in homes and businesses have crowded the 2.4 GHz ISM band which is shared by both Wireless Sensor Network (WSN) and WiFi. This coexistence has been shown to be prone to lot of interference \cite{ref:intf} from other WiFi networks, Bluetooth and Microwaves. As interference is a decisive factor in a radio packet transmission, it manifests in sizable issues like detecting if a packet is the intended packet or not [8], and deciphering the incoming packet. As a result, it leads to unnecessary awakes of radio, repeat transmissions, and unrecoverable packets. To circumvent this, many approaches have been suggested from smarter algorithms detecting interference to making radios efficient. However, the use of a low band did not seem to gather much traction in the WSN field. We chose to study a 868 MHz based wireless sensor network and our contributions henceforth is as follows

\begin{itemize}
\item We setup a 868 MHz wireless sensor network in an university environment to study the signal strength, communication range, and packet loss rate of the radio.
\item We conduct experiments to assess the feasibility of long-range communication with 868 MHz based on this wireless sensor network.
\item We show that our deployed network of 868 MHz can achieve satisfactory coverage in large area with small number of nodes.
\end{itemize}


 



%
%


\section{Background}

WSN is typically a low power sensor network monitoring a parameter of interest \cite{survey:overview} that could be anything ranging from temperature of a room to a heartbeat of a person. WSN has manifested itself into Internet of Things (IoT), which predicts the connectedness \cite{adhoc:868model} of things. Figure 1 depicts spatially distributed sensor nodes which collect parameter(s) of interest and send it to a base station which further relays it to Internet. 

\hfill \linebreak
\includegraphics[scale=0.41]{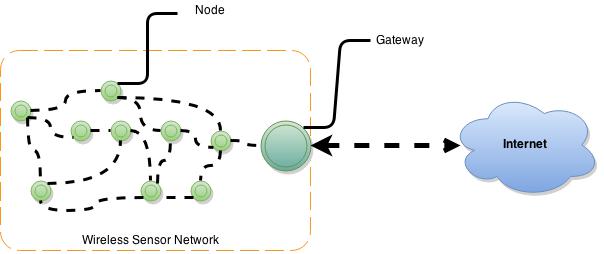}
\centerline{Figure 1 - Wireless Sensor Network}
\hfill

Various types of sensor networks exists on different protocols namely ZigBee, 802.15.4 and 6LowPAN. The frequency band at 2.4GHz in which most of the WSN operate is prone to multi-path fading, and known to expose the nodes to interference from other technologies like Bluetooth, WiFi or micro wave ovens. Studies comparing the IEEE 802.15.4 link performance with meteorological conditions have shown a high correlation between the packet reception rate and temperature \cite{ltermmetstudy:uu}, with better performance at lower temperatures. In this paper, we will explore long-range wireless communication with 868 MHz and study its performance in outdoor environment.

%
%

\section{WSN Setup with 868 MHz radio}

We developed the sensor nodes ourselves integrated with 868 MHz radio antenna for this wireless sensor network. Our nodes were built on Seediuno Stalker board which houses ATMega328P micro controller with Arduino \cite{arduino:linka} and XBee support. The 868 MHz radios were used from Digi International. 

The nodes were deployed and spread out between the classrooms in different buildings. Each node was equipped with temperature, humidity, oxygen and particulate matter sensor. The payload was 100 bytes comprised of output of the aforementioned sensors, local time, and analog readings of 7 pins. All the nodes were given constant supply of 5V. Test runs were done with batteries and solar plugins, but the final setup was with plugged-in wall chargers.

pcDuino \cite{pcduino:link} was chosen for the base station. It features an on-board computer platform running Lubuntu OS, and has built-in support for Arduino. Nodes were loaded with same Arduino based sketches, and the base station running a Python program was used to capture the incoming frames from nodes. The frames were further time stamped in the base station and pushed to an IoT cloud platform for aggregation and visualization. 
\hfill \linebreak
Our WSN deployment includes the following nodes.

Base Station - Placed in Building 1, 4th floor approximately 30m from ground level, and 5m from windows.

Node 001 - Placed indoors and within close proximity of the base station.

Node 002 - Placed in adjacent building, approximately 80m away, and 50m from ground level.

Node 003 - Placed in the same building, approximately 50m away, and 50m from ground level.

The buildings are historic, and have 20cm thick brick walls with heavy use of metal for ventilation inside.  

The maximum distance achieved indoors from the base station was 280 meters. We did some test trials at these distances before permanently placing them at these locations. Figure 2 shows the placement of nodes with respect to base station. 
\hfill \linebreak

\includegraphics[scale=0.50]{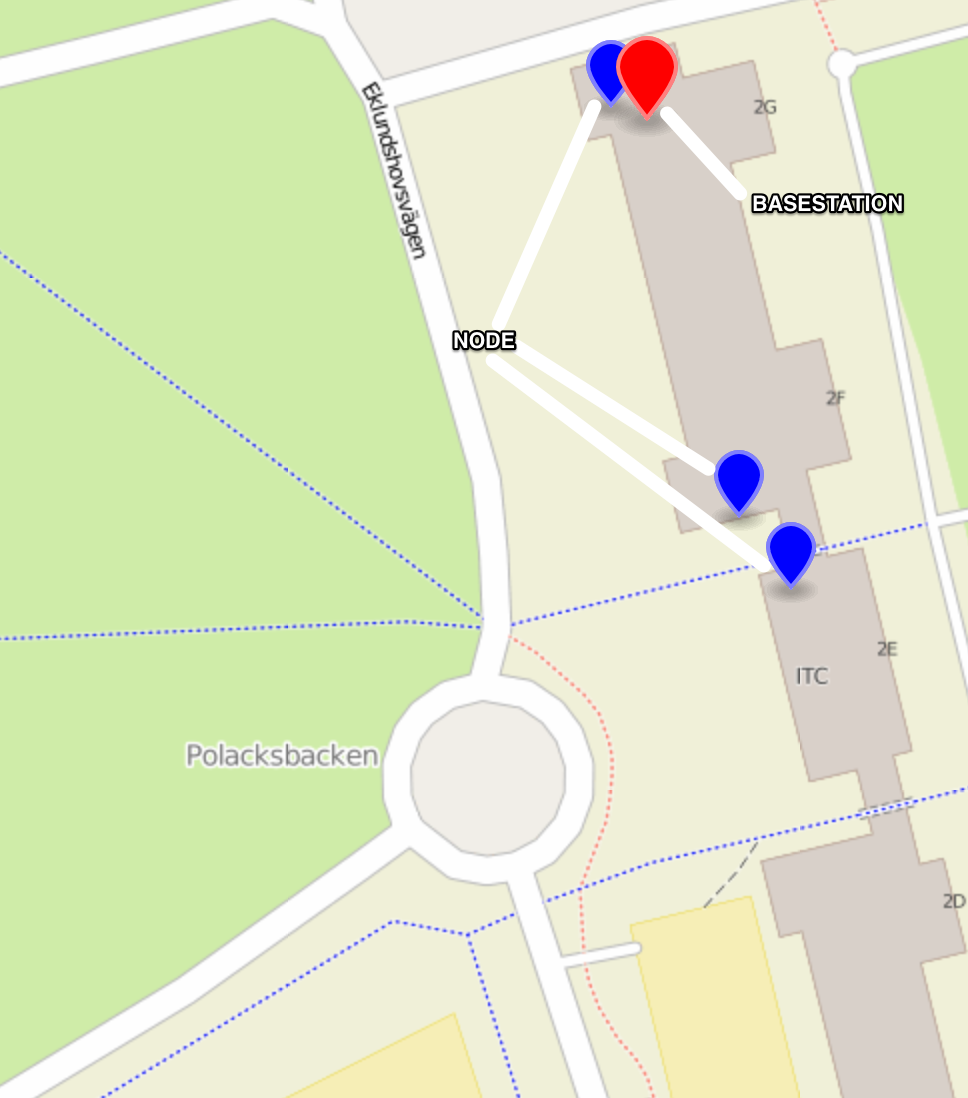}
\centerline{Figure 2 - Deployed Sensor Network}

%
\section{Experiments}

\subsection{Experiment settings}

The Nodes were connected to USB power, and were placed in one classroom, and two office rooms. The base station and nodes were placed across two buildings adjacent to each other. The readings were taken over a period of 1 week where temperature, particulate matter (PM), humidity and oxygen was measured at a 3 minute interval.

No duty cycle was programmed for the nodes, the radios were always on. The order of sampling was temperature, oxygen, humidity, and PM. The packet was then broadcasted, the node radios were all programmed to be coordinators, hence a point to point link network was being used.

The nodes were  manually synced to 30 seconds apart, these nodes have internal RTC. The base station was set to be awake all the time. The sampled time is appended by the base station as it parses the incoming packets.

The transmitting power on the nodes was set to 300mW (24dBm), which is programmable in XBee. This is one of the reasons we adopted this radio.

We compared the three nodes to measure the Packet Error Rate (PER). We define PER as the number of packets arrive in an hour. There will be 20 packets in an hour given that each node sent at an interval of 3 minutes.


\subsection{Results}

The experiment was conducted for a week and the packet error rate evaluated in sensor data collection. The inference of packet loss is determined by looking at the arriving packets at the base station, where non-arrival means packet loss. We analyzed the 24 hour readings from Node 001 and 002. Figure 3 shows a snapshot of a 24 hour window and no packet loss was observed in Nodes 001 and 002. Node 003 showed similar behavior as Node 001. 
\hfill \linebreak

\includegraphics[scale=0.34]{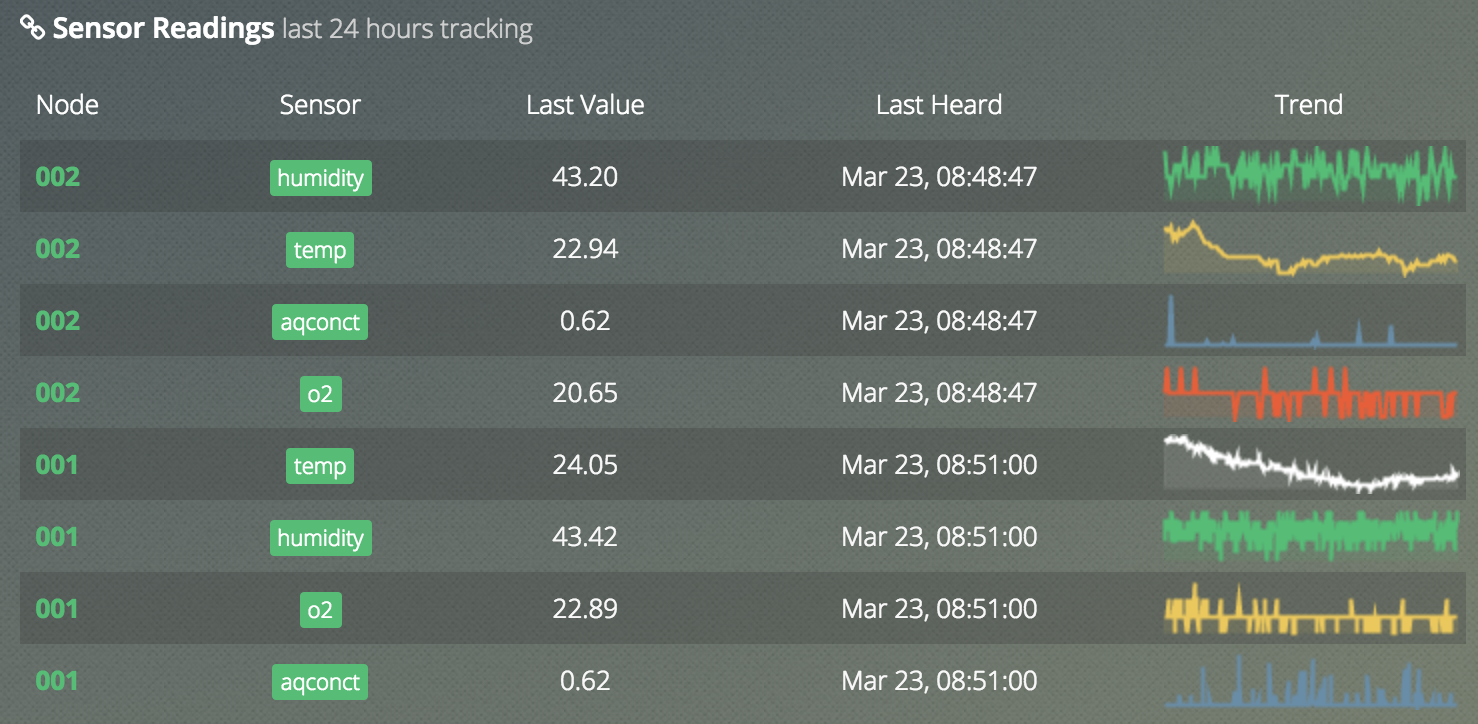}
\centerline{Figure 3 - Snapshot 24Hrs reading}

\hfill

However, during the week of measurements, significant packet loss was observed in other time periods. In Figures 4 and 5, it is shown that during another 24 hour period, Node 002 experienced packet loss whereas there was no packet loss for Node 001 (All data points present every 3 minutes).

\includegraphics[scale=0.60]{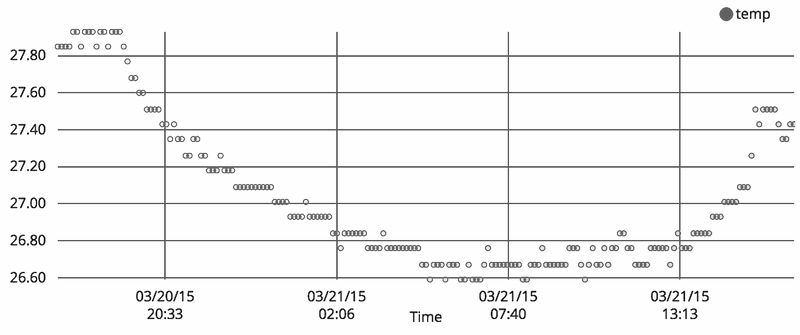}
\centerline{Figure 4 - 24Hrs Node 001}

\includegraphics[scale=0.60]{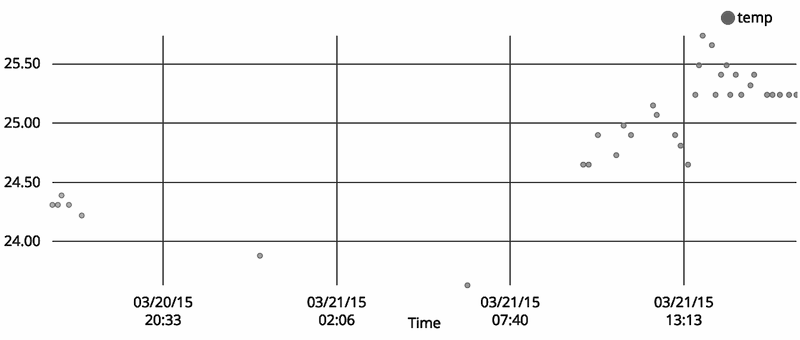}
\centerline{Figure 5 - 24Hrs Node 002}
As a reminder, Node 001 is collocated with the base station, and Node 002 is 80m away in a detached adjacent building. As one can see Node 002 has sizable packet loss which provokes curiosity.

We briefly postulate possible causes for packet loss. It can be assumed that as nodes are plugged into the wall, and noise floor for 868 MHz is low \cite{credit:868}, we can rule out external interference i.e. interference from 2.4 GHz networks and Microwaves.Next, two plausible candidates to consider was \emph{environmental factors} and \emph{fading}. 

We first study the environmental parameters between the buildings as they are detached, and we looked into the prevalent conditions during the times when we experienced heavy packet loss. We were able to exclude the indoor climate conditions around the nodes as the ambient temperature was controlled, (Figures 4 and 5) exhibit the relatively minuscule variation of temperature, this helped us shift our focus to outdoors.To explore this point, we got the local weather station data giving us the Temperature, Dew point and Atmospheric pressure (see Figure 6 and 7, Courtesy Weather Underground). There seems to be some correlation between environmental conditions and packet loss which needs to be studied further. Additionally, its supplemented by the fact that troposphere consists of different kind of particles and hydrometeors, and have various degrees of effect on radio propagation \cite{book1:rpatwc}. 
\par{For the current scenario we limit our observation to rain drops, snow and fog as these were the common occurrences during our study. Readings from the local weather station for the day of interest is mentioned just as a reference for preliminary consideration, further research has to be carried out to make use of the local micro climate for more meaningful analysis and conclusion.}

Our hypothesis is further supported by \cite{book1:rpatwc}, as it credits wave loss in the aforementioned conditions to be the combined loss due to absorption and scattering, that is to say as below

Per \cite{book1:rpatwc} Total wave loss(dB) \[ L_{tot} = L_{abs} + L_{scat}  \]

In the above equation, \emph{L\textsubscript{abs}} is the loss in decibel due to \emph{Absorption}, and \emph{L\textsubscript{scat}} is the loss in decibel due to \emph{Scattering}. \emph{Absorption} is in effect conversion of radio wave energy to thermal energy, and \emph{Scattering} is redirection of radio waves into various directions\cite{book1:rpatwc}. As indicated before, additional analysis needs to be carried out to ascertain its effect or if otherwise.  

\includegraphics[scale=0.35]{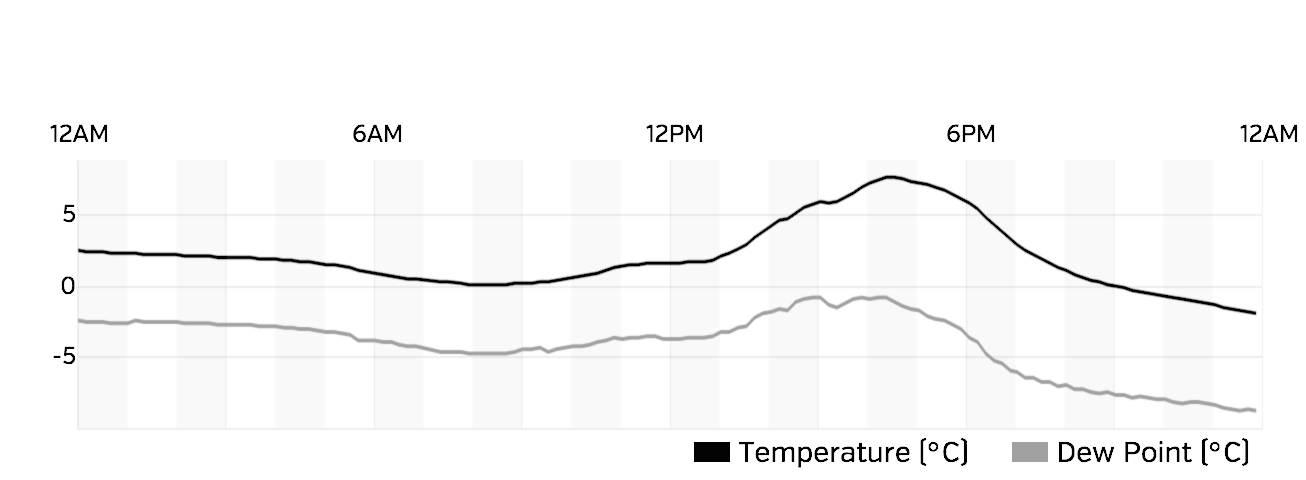}
\centerline{Figure 6 - Temp \& Dew point}

\includegraphics[scale=0.35]{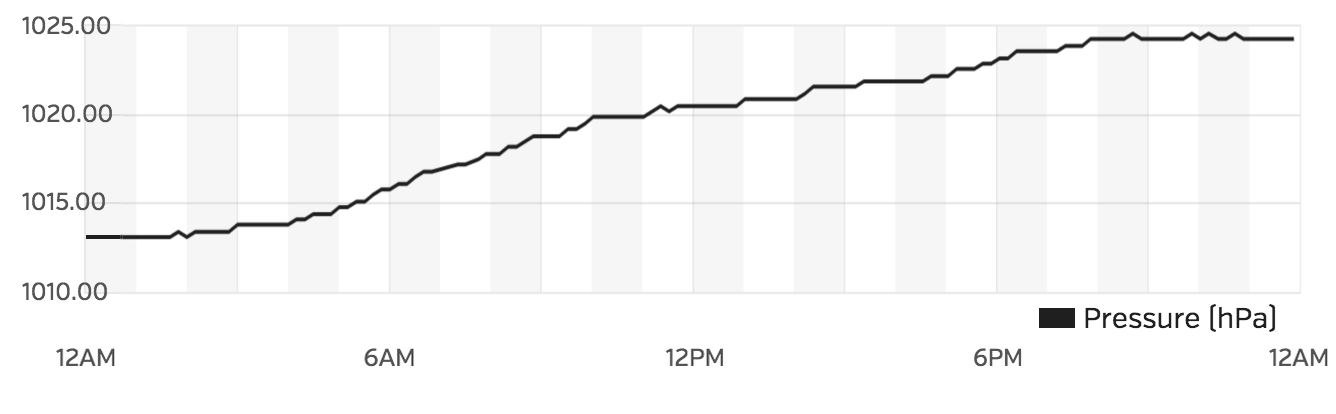}
\centerline{Figure 7 - Pressure}



Another possible cause of packet loss is Fading. Here we bring it up as a possibility as the antennas were not directionally positioned, and could be playing a part in weaker signal at the base station leading to packet losses \cite{fading:a868}. 

\subsection{Outdoor Transmission Performance}
For the last part of our study, we looked at the Line-Of-Sight (LOS) coverage of our 868 MHz WSN. Figure 8 shows the heat map depicting the coverage, where lighter color indicates lower RSSI strength. It showed positive results for range up to 300m, which is much further than the existing protocols in 2.4 GHz. Positive result was inferred, i.e. average RSSI strength above 40 percent at all locations with no packet loss observed.
\hfill \linebreak

\includegraphics[scale=0.45]{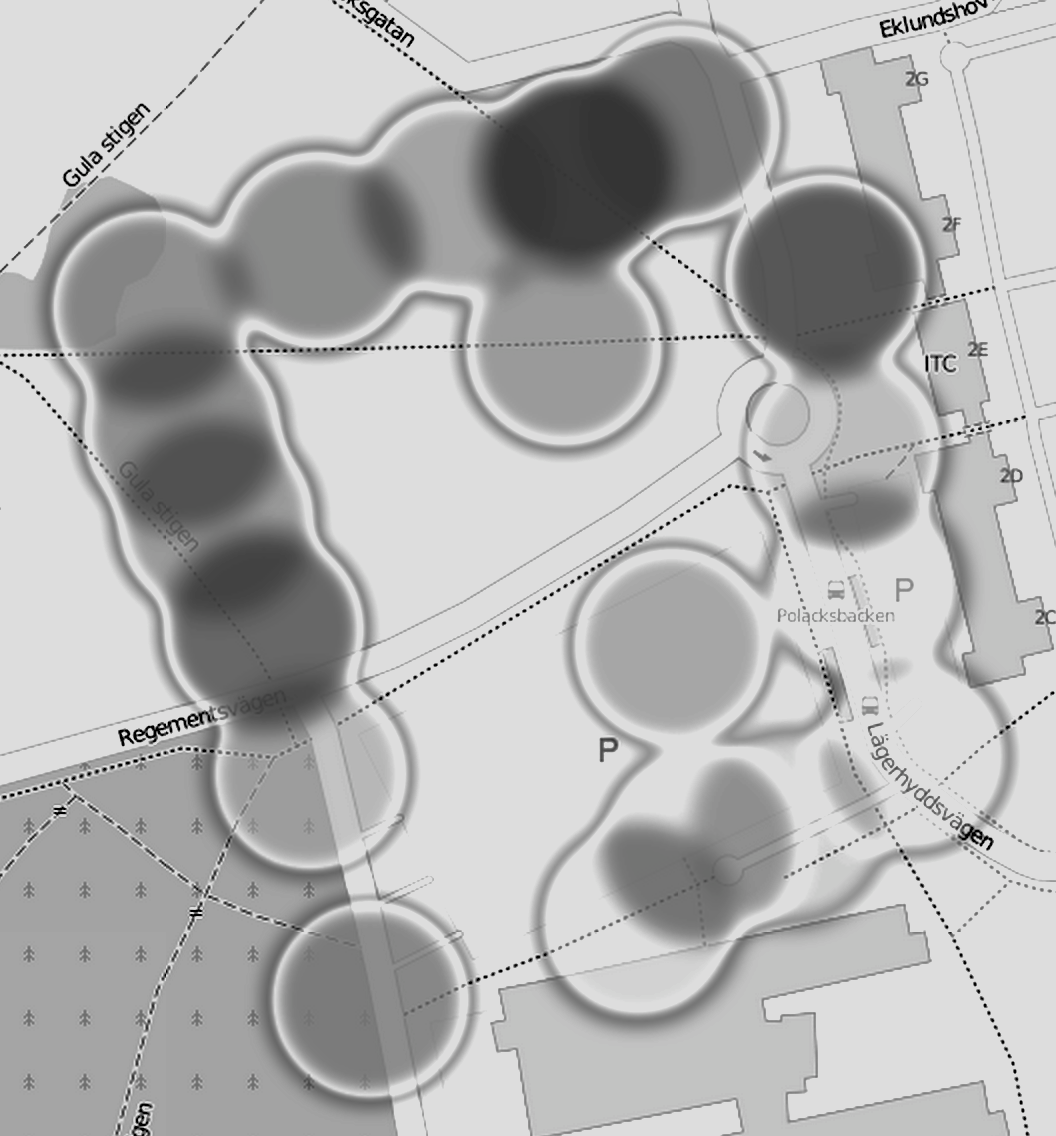}
\centerline{Figure 8 - LOS Coverage 868 MHz}

\section{Conclusions and Future Work}
In conclusion, we deployed a wireless sensor network with 868 MHz radio and conducted a case study to evaluate the performance of the radio. The results were promising, but the packet losses were intriguing. Of the three nodes only one showed significant packet loss. Since this is the only node located in another building, we hypothesize that the air pocket between the buildings is affecting transmission and weather changes (high humidity, rain etc.) are causing packet loss, or weakening radio signal by fading. This will be investigated in future work. A study conducted by Uppsala University researchers concludes that only temperature is a significant factor, higher temperature causing packet loss \cite{ltermmetstudy:uu}. However, this was tested using 2.4 GHz radios in outdoor conditions, so we could perform a similar study using 868 MHz radios. The method of testing would be similar to the one used before: two nodes equidistant to the base station, only one having an air pocket in between, outdoor weather conditions (humidity, temperature etc.) would also be measured.
\par{A way to extend transmission range/ameliorate fading would be to add a high-gain antenna or adding more nodes (relays) in the network. However, packet loss was not observed within a 300m distance and this transmission distance is enough with a sufficient amount of (relay) nodes.}
\par{In future work, we would next look into deploying this technology in a smart city monitoring program and habitat monitoring. Due to the large transmission range outdoors, it would be possible to deploy mobile nodes that detect a base-station within a 300m range and dump the data collected, which then gets forwarded to the Cloud. Using this approach, mobile nodes could be used on buses to measure air quality and obtain pollution heat-map of any burgeoning city. Additionally, we plan to investigate 4G/LTE User Equipments(UE) interference on 868MHz as studied by \cite{4glteinterf:ue}.}


\section*{Acknowledgment}
The project has been supported by the VINNOVA GreenIoT project for smart cities.  

\newpage


\end{document}